\documentstyle[12pt,epsfig]{article}
\date{}
\begin{document}
\noindent
\makebox[140mm][r]{NPI 97-1}
\rule{140mm}{0.5mm}
\rule[4.5mm]{140mm}{1mm}
\vspace{35mm} 
\begin{center}
{\Large\bf
NEUTRAL PION ELECTROPRODUCTION ON THE PROTON NEAR THRESHOLD     
}
\end{center}
\begin{center}
\vspace{0.5cm}
{\large\sl
                      A.N.Tabachenko \\ 
}
\vspace{0.5cm}
{\it
      Nuclear Physics Institute , 634050 Tomsk, P.O.B.25, RUSSIA\\
      email:ant@npi.tpu.ru
              
}

\end{center}
\begin{abstract}
The neutral pion electroproduction on the proton is considered in the
framework phenomenological model based on the PCAC, current algebra with
hard pions and chiral symmetry breaking interaction. The s- wave
differential cross-section of the electroproduction of neutral pions off
protons near the threshold is predicted as function photon momentum transfer
squared $k^2$ . The s- wave multipoles $ReE_{0+}$ and $ReL_{0+}$ as function
of the energy of the $\pi^0p$- system are given for the small $k^2$ and $%
k^2=0.1(GeV/c)^2$. The results are compered with the experiment. It is shown
the significant role of the explicit chiral symmetry breaking.
\end{abstract}

\vspace{1.5cm}

\begin{flushright}
{\large\rm JUNE 27, 1997\\
}
\end{flushright}
\noindent
\rule{140mm}{1mm}
\rule[5mm]{140mm}{0.3mm}

\newpage

\noindent{\bf 1. Introduction} \vspace{2mm}

The processes of the pion photo-electroproduction off the nucleon near the
threshold are of interest to the examination of the such fundamental
principles as gauge invariance, partial conservation of the axial-vector
current (PCAC). These principles together with the requirement of the
analyticity and cross-symmetry allow to receive the model-independent
prediction for the amplitude on the threshold, so called low- energy theorem
(LET). The predictions of the LET were confronted with new accurate
experimental dates. The analysis of the accurate experimental data at Mainz
[1] about the differential and total cross-sections of the neutral pion
photoproducion off proton near threshold was made in the works [2, 3]. The
electric dipole amplitude $E_{0+}^{}$ at the threshold was founded using
additional assumption concerning the p-waves near threshold. The value of
the amplitude $E_{0+}=(-2.0\pm 0.2)*10^{-3}/m_{\pi ^{+}}$ turns out in
quantitatively agreement with the prediction of the LET $%
E_{0+}=-2.3*10^{-3}/m_{\pi ^{+}}$. This analysis gives also the rapid energy
variation of the s-wave $E_{0+}$ amplitude when the energy $\pi N$ -system
varied from the $\pi ^0p$ threshold to the $\pi ^{+}n$ threshold. However,
the new neutral pion photoproduction off proton data from the TAPS
collaboration [4] and SAL [5] show same discrepancies to the previously
considered best data of Beck at.al. The new results give also the rapid
energy variation of the s-wave $E_{0+}$ amplitude between the $\pi ^0p$ and
the $\pi ^{+}n$ thresholds, but the values of the E$_{0+}$ on the $\pi ^0p$
and the $\pi ^{+}n$ thresholds are different from the result [2,3] and LET.
The experimental values of the $E_{0+}$ from TAPS data are $E_{0+}^{exp}(\pi
^0p)=(-1.31\pm 0.08)*10^{-3}/m_{\pi ^{+}}$ and $E_{0+}^{exp}(\pi
^{+}n)=-0.4*10^{-3}/m_{\pi {+}}$.

The discrepancies between the new data of TAPS and LET is caused possibly by
the incomplete LET. As it was shown in the baryon chiral perturbation theory
(CHPT) [6] described by an effective Lagrangian incorporating global
symmetry of QCD the large chiral loop corrections are needed for the s-wave
multipole on the threshold [7,8]. The s-wave amplitudes $E_{0+}$ in CHPT
have the rapid energy behavior between the $\pi ^0p$ and the $\pi ^{+}n$
thresholds and are $E_{0+}(\pi ^0p)=-1.16*10^{-3}/m_{\pi ^{+}}$ and $%
E_{0+}(\pi ^{+}n)=-0.44*10^{-3}/m_{\pi {+}}$ respectively. This results are
close to the new experimental values of the $E_{0+}$ from TAPS data.

The rapid energy variation of the s-wave $E_{0+}$ amplitude was received
also in the phenomenological model [9] which was based on the above
mentioned principles of the gauge invariance and PCAC
and the ''Cloudy-Bag'' model. In this approach the
amplitude of the process contained the part received with help current
algebra with hard pions and the additional part connected with a symmetry
breaking term analogous to the $"Sigma-term"$ in the low-energy pion-nucleon
scatter. The corresponding values s-waves $E_{0+}$ on the $\pi ^0p$ and the $%
\pi ^{+}n$ thresholds are $E_{0+}(\pi ^0p)=-1.50*10^{-3}/m_{\pi ^{+}}$ and $%
E_{0+}(\pi ^{+}n)=-0.50*10^{-3}$.

By studying the pion electroproduction off nucleon we are able to consider
the same problems connected with $E_{0+}$ multipole when it depends from the
4-momentum squared of the virtual photon $k^2$. In addition, due to the
longitudinal coupling of the virtual photon to the nucleon spin it is
possible to consider the s- wave longitudinal multipole $L_{0+}$ and its
dependence from $k^2$. At last, in the pion electroproduction on the proton
four structure function instead one in the photoproduction (for unpolarized
particles) allow to do more detailed a test of the prediction of the various
theoretical models.

Over the last years there has been considerable activity to precisely
measure neutral pion electroproduction in the threshold region very close to
threshold, small photon momentum transfer squared $k^2$ and typical photon
momentum transfer squared $k^2=-0.1(GeV/c)^2$. In the work of Welch et al.
[10] at NIKHEF the s-wave cross-section was measured. The kinematically
complete differential cross-sections for various photon polarization $%
\epsilon =0.5 ,..., 0.9$ were given by Van den Brink et al. [11] (NIKHEF) and
Distler et al. [12] (MAMI). The data at lower $|k^2|$ have also been taken
at MAMI [13].

Recently, the systematic analysis of the processes of the pion
electroproduction on the nucleon in the threshold region was made in the
baryon chiral perturbation theory [14]. Prediction for the lower photon
virtualities for the s- and-p waves multipoles and differential cross
sections were given.

In this work we use the generalized phenomenological approach [9] for the
research of the neutral pion electroproduction on the proton near threshold.
We consider the s- wave dipole electric and longitudinal multipoles and
differential cross sections as function of the photon momentum transfer
squared $k^2$ and the energy $\pi N$-system near threshold.

\begin{center}
2. Pion Electroproduction Amplitude.
\end{center}

Let us consider amplitude $S_{a\nu}$ which is connected to the matrix
element of the S-matrix of the electroproduction of pions off nucleons with
the help of the relation

$$
<f|S|i>=-(2\pi )^4\delta (p_1+k-p_2-q)N_kN_qS_{a\nu }\epsilon ^\nu (\vec k). 
$$
Here $N_k,N_q$ are the normalization factors of the photon and
pion, respectively; $\epsilon ^\nu (\vec k)$ is the polarization of the 
virtual photon; $p_1,p_2$ are the
four-momenta of the incoming and outgoing nucleons, respectively. Using the
partially conserved axial-vector current (PCAC) hypothesis, one can write [9]

$$
S_{a\nu}={\frac{1}{{F_\pi}}}\{iq^{\mu}T_{a\mu\nu}-C_{a0\nu}-{\frac{iq^0}{{\
m_{\pi}^2}}}\Sigma_{a\nu}\}.\eqno(1)%
$$

Here 
$$
T_{a\mu\nu}=\int dx e^{-iqx} < p_{2} \vert T(\tilde{A}_{a\mu}(x)J_{\nu} (0))
\vert p_{1}>\eqno(1a)%
$$

is the axial-vector amplitude,

$$
C_{a\mu\nu}=\int dx e^{-iqx} \delta (x_0)<p_{2}\vert [A_{a\mu}(x),j_{\nu}
(0)]\vert p_{1}> \eqno(1b)%
$$
and

$$
\Sigma_{a\nu}=\int dx e^{-iqx} \delta(x_0)<p_{2}\vert [{\frac{\partial }{{\
\partial x_{\mu}}}}A_{a\mu}(x),j_{\mu}(0)]\vert p_{1}> \eqno(1c)%
$$

are connected with the matrix elements of the commutators between the axial-
vector current $A_{a\mu}(x)$ and electromagnetic current $J_{\nu}(0)$ and
the divergence of the axial-vector current and electromagnetic
current, respectively. The value $\tilde{A}_{a\mu}(x)$ is given by 
$$
\tilde {A}_{a\mu}(x)=A_{a\mu}(x)-F_{\pi}{\frac{\partial}{{\partial x_ {\mu}}}%
}\pi_{a}(x),%
$$

where $\pi _a(x)$ is an interpolating field for the pion. For the axial
vector amplitude $T_{a\mu \nu }$, the conservation of the electromagnetic
current gives the condition

$$
ik^{\nu}T_{a\mu\nu}=C_{a\mu0}+{\frac{i(k_{\nu}\delta_{\mu0}-q_{\mu}\delta_{%
\nu0}) }{{m^{2}_{\pi}}}}\Sigma_{a\nu}.\eqno(2)%
$$

Now using the method which was worked out in the Ref. [15], one separates the
contribution for the single- particle intermediate states from the values
(1a-1c). We define the nonpole part of the $iq^\mu T_{a\mu \nu }$ as 
$$
iq^\mu T_{a\mu \nu }=iF_\pi \sum_{B,\alpha _B}(\frac 1{N_{p_B}^2}\frac{%
<p_2\mid j_{\pi a}(0)\mid p_B\alpha _B><p_B\alpha _B\mid j_\nu (0)\mid p_1>}{%
s-m_B^2} 
$$
$$
+\frac 1{N_{p_B^{^{\prime }}}^2}\frac{<p_2\mid j_{\pi a}(0)\mid
p_B^{^{\prime }}\alpha _B><p_B^{^{\prime }}\alpha _B\mid j_\nu (0)\mid p_1>}{%
u-m_B^2}) 
$$
$$
-\sum_{M,\alpha _M}iq^\mu T_{a\mu \nu }(M)\frac 1{t-m_M^2}\frac{N_{p_2}}{%
N_{p_M}^2}\overline{u}(\overrightarrow{p}_2)\times <p_M\alpha _M\mid \eta
(0)\mid p_1>+iq^\mu \widetilde{T}_{a\mu \nu .}\eqno(3) 
$$

Here B-index of the single-particle intermediate states with mass $M_{B}$
in s- and u- channels; M-index of mesons in t- channels; $\alpha
_B,\alpha _M$ define the isotopic and spin states; $\eta (0)$ is defined as 
$(i\gamma ^\mu \partial _\mu +M)\psi (x)=\eta (x)$, where $\psi (x)$ is the
field of the barion; the value $T_{a\mu \nu }(M)$ is defined as

$$
\begin{array}{c}
T_{a\mu \nu }(M)=\int d^4ye^{-iqy}<0\mid T(
\widetilde{A}_{a\mu \nu }(y)j_\nu (0))\mid p_M\alpha _M>; \\ 
\end{array}
$$
$iq^\mu \widetilde{T}_{a\mu \nu }$ is the nonpole-part of the amplitude $%
iq^\mu T_{a\mu \nu }$. The $iq^\mu \widetilde{T}_{a\mu \nu }$ includes
contact terms and the contributions from the many-partical intermediate
states. Analogically, we have

$$
C_{a0\nu }(\Sigma _{a\nu })=-\sum_{M,\alpha _M}C_{a0\nu }(\Sigma _{a\nu
})(M)\frac 1{t-m_M^2}\frac{N_{p_2}}{N_{P_M}^2}\overline{u}(\overrightarrow{p}%
_2)<p_M\alpha _M\mid \eta (0)\mid p_{1>} 
$$
$$
+\widetilde{C}_{a0\nu }(\widetilde{\Sigma }_{a\nu }),\eqno(4) 
$$
where 
$$
\begin{array}{c}
C_{a0\nu }(M)=\int dxe^{-iqx}\delta (x_0)<p_M\alpha _M\mid [A_{a0}(x),j_\nu
(0)]\mid p_1>, \\  
\\ 
\Sigma _{a\nu }(M)=\int dxe^{-iqx}\delta (x_0)<p_M\alpha _M\mid [\frac
\partial {\partial x_\mu }A_{a\mu }(x).j_\nu (0)]\mid p_1>. 
\end{array}
$$
The $\widetilde{C}_{a0\nu }(\widetilde{\Sigma }a\nu )$ are the non-pole parts
of the $C_{a0\nu }(\Sigma _{a\nu })$. Now we use the equation analogical to
the equation (1) 
$$
S_{a\nu }(M)=\frac 1{F_\pi }\{iq^\mu T_{a\mu \nu }(M)-C{a\nu }(M)-\frac{iq^0%
}{m_\pi ^2}\Sigma _{a\nu }(M)\},\eqno(5) 
$$
where $S_{a\nu }(M)$ is defined as 
$$
\begin{array}{c}
<\pi qa\mid S\mid p_M\alpha _M;\gamma k\varepsilon >=-(2\pi )^4\delta
(p_M+k-q)N_kN_qS_{a\nu }(M)\varepsilon ^\nu ( 
\overrightarrow{k}). \\  
\end{array}
$$
This yields

$$
S_{a\nu} = S_{a\nu}(P) +{\frac{1}{{F_{\pi}}}}\{iq^{\mu}\tilde R_{a\mu\nu}
-\tilde C_{a0\nu}\}\eqno(6) 
$$

and 
$$
ik^{\nu}\tilde R_{a\mu\nu} = T_{a\mu}(P) +\tilde C_{a\mu0} \eqno(7) 
$$

instead of Eqs.(1) and (2). Here

$$
\tilde R_{a\mu \nu }=\tilde T_{a\mu \nu }-{\frac{\delta _{\mu 0}}{{m_\pi ^2}}%
}\tilde \Sigma _{a\nu }\eqno(8) 
$$

and $\sim $ means that the contributions from the single-piratical
intermediate states are removed from the considered values. The $S_{a\nu }(P)$
is the sum of the contributions from the single-particle intermediate states
in the s-, u- and t- channels

$$
S_{a\nu }(P)=i\sum_B\bigl[N_{p_B}^{-2}{\frac{{<p_2|j_{{\pi }%
a}(0)|p_B><p_B|j_\nu (0)|p_1>}}{{s-m_B^2}}} 
$$
$$
+N_{p_B^{\prime }}^{-2}{\frac{{<p_2|j_\nu (0)|p_B^{\prime }><p_B^{\prime
}|j_{{\pi }a}(0)|p_1>}}{{u-m_B^2}}}\bigr] 
$$
$$
+i\sum_MN_{p_M}^{-2}N_{p_2}N_q^{-1}{\frac{{\bar u(\overrightarrow{p}_2)<\pi qa|j_\nu
(0)|p_M><p_M|\eta (0)|p_1>}}{{t-m_M^2}},}\eqno(9) 
$$

and the $T_{a\mu }(P)$ is an expression of the form

$$
T_{a\mu}(P)=i\sum_{B}\bigl[N^{-2}_{p_{B}}{\frac{{<p_{2}\vert \tilde
A_{a\mu}(0)\vert p_{B} ><p_{B}\vert -ik^{\nu}j_{\nu}(0)\vert p_{1}>}}{{%
s-m_{B}^2}}}%
$$
$$
+N_{p^{\prime}_{B}}^{-2}{\frac{{<p_{2}\vert-ik^{\nu} j_{\nu}(0)\vert
p^{\prime}_{B}>< p^{\prime}_{B}\vert \tilde A_{a\mu}(0)\vert p_{1}>}}{{%
u-m_{B}^{2}}}}\bigr].\eqno(10)%
$$

For the amplitude $iq^\mu \tilde R_{a\mu \nu }$, one has the condition

$$
k^{\nu}(iq^{\mu}\tilde R_{a\mu\nu})=q^{\mu}T_{a\mu}(P)+q^{\mu}\tilde
C_{a\mu0}.\eqno(11)%
$$

Consider the linear independent basis

$$
O_{i=1...6},N_{4\nu},N_{7\nu},\eqno(12)%
$$

where $O_i$ are defined by Adler [16] as

$$
O_{1\nu}={\frac{1}{2}}i\gamma_{5}[\gamma_{\nu}\hat k-\hat k\gamma_{\nu}]%
$$

$$
O_{2\nu}=2i\gamma_{5}[qkP_{\nu}-Pkq_{\nu}]%
$$

$$
O_{3\nu}=2\gamma_{5}[qk\gamma_{\nu}-\hat kq_{\nu}]%
$$

$$
O_{4\nu}=2\gamma_{5}[Pk\gamma_{\nu}-\hat kP_{\nu}]%
$$

$$
O_{5\nu}=i\gamma_{5}[qkk_{\nu}- k^{2}q_{\nu}]%
$$

$$
O_{6\nu}=\gamma_{5}[\hat kk_{\nu}-k^{2}\gamma_{\nu}]%
$$

and $N_i$ are defined by Ball [17] as

$$
N_{4\nu}=2i\gamma_{5}k_{\nu}%
$$

$$
N_{7\nu}=\gamma_{5}\hat kk_{\nu}.%
$$

Let us expand the value ${\frac{{i}}{{F_\pi }}}q^\mu \tilde R_{a\mu \nu }$
in this basis. Then from equation (11) we have for the value connected with
the structures $N_{4\nu }$ and $N_{7\nu }$ exprassion

$$
\bigl( \frac i{F_\pi }q^\mu \tilde R_{a\mu \nu }\bigr)%
^{NGI}=-iN_{p_1}N_{p_2}(\frac{\alpha _a}{k^2}\overline{u}(
\overrightarrow{p}_2)N_{4\nu
}u(\overrightarrow{p}_1)+\frac{\beta _a}{k^2}
\overline{u}(\overrightarrow{p}_2)N_{7\nu }u(\overrightarrow{p}_1)),\eqno(13) 
$$

where $\alpha _a$ and $\beta _a$ are the functions which are defined by the
right part of the equation (11); $NGI$ denotes the gauge noninvariant part
of the amplitude in the mentioned basis above. Let us govern also the value

$$
\bigl({\frac{{i}}{{F_\pi }}}q^\mu \tilde R_{a\mu \nu }\bigr)^{GI}={\frac{{i}%
}{{F_\pi }}}q^\mu \tilde R_{a\mu \nu }-\bigl({\frac{{i}}{{F_\pi }}}q^\mu
\tilde R_{a\mu \nu }\bigr)^{NGI},\eqno(14) 
$$
It is obvious, that $\bigl({\frac{{i}}{{F_\pi }}}q^\mu \tilde R_{a\mu \nu }%
\bigr)^{GI}$ satisfied the equation

$$
k_\nu \bigl({\frac{{i}}{{F_\pi }}}q^\mu \tilde R_{a\mu \nu }\bigr)^{GI}=0,\eqno(15) 
$$
 i.e. it is the gauge invariant value. Let us write

$$
{\frac{{i}}{{F_{\pi}}}}q^{\mu}\tilde T_{a\mu\nu}= \bigl( {\frac{i}{{F_{\pi}}}%
}q^{\mu}\tilde T_{a\mu\nu}\bigr)^{GI}+ X_{a7}\overline{u}(\overrightarrow{p}_
{2})N_{4\nu}u(\overrightarrow{p}_{1})+X_{a8}\overline{u}(\overrightarrow{p}_
{2})N_{7\nu}u(\overrightarrow{p}_{1})%
$$

$$
-i{\frac{q_{\mu}\delta_{\mu0} }{{F_{\pi}m^{2}_{\pi}}}}\tilde \Sigma_{a\nu}= 
\bigl(-i{\frac{q_{\mu}\delta_{\mu0} }{{F_{\pi}m^{2}_{\pi}}}}\tilde
\Sigma_{a\nu}\bigr)^{GI} +Y_{a7}\overline{u}(\overrightarrow{p}_{2})
N_{4\nu}u(\overrightarrow{p}_{1})+Y_{a8}%
\overline{u}(\overrightarrow{p}_{2})N_{7\nu}u(\overrightarrow{p}_{1}),%
$$
were

$$
\bigl( {\frac{i}{{F_{\pi}}}}q^{\mu}\tilde T_{a\mu\nu}\bigr)^{GI}=
\sum_{i=1}^{6}
X_{ai}\overline{u}(\overrightarrow{p}_{2})O_{i\nu}u(\overrightarrow{p}_{1})%
$$
and

$$
\bigl(-i{\frac{q_{\mu}\delta_{\mu0} }{{F_{\pi}m^{2}_{\pi}}}}\tilde
\Sigma_{a\nu}\bigr)^{GI}= \sum_{i=1}^{6}Y_{ai}\overline{u}%
(\overrightarrow{p}_{2})O_{i\nu}u(\overrightarrow{p}_{1}).%
$$

Then the equation (11) imposes the condition on the $X_{a7,8},Y_{a7,8}$

$$
X_{a7}+Y_{a7}={\frac{\alpha_{a}}{k^{2}}}%
$$
$$
\eqno(16)%
$$
$$
X_{a8}+Y_{a8}={\frac{\beta_{a}}{k^{2}}}.%
$$

As a result, we have for the gauge-invariant value $\bigl({\frac{{i}}{{F_\pi 
}}}q^\mu \tilde R_{a\mu \nu }\bigr)^{GI}$ the expression

$$
\bigl({\frac{{i}}{{F_{\pi}}}}q^{\mu}\tilde R_{a\mu\nu}\bigr)^{GI}= \bigl( {%
\frac{i}{{F_{\pi}}}}q^{\mu}\tilde T_{a\mu\nu}\bigr)^{GI}+ \bigl(-i{\frac{%
q_{\mu}\delta_{\mu0} }{{F_{\pi}m^{2}_{\pi}}}}\tilde \Sigma_{a\nu}\bigr)^{GI}.%
\eqno(17)%
$$
Then the amplitude of the electroproduction of pions off nucleons can be
written as

$$
S_{a\nu}=S_{a\nu}^0 +\bigl({\frac{{i}}{{F_{\pi}}}}q^{\mu} \tilde R_{a\mu\nu}%
\bigr)^{GI},\eqno(18)%
$$
where

$$
S_{a\nu }^0=S_{a\nu }(P)+\bigl({\frac{{i}}{{F_\pi }}}q^\mu \tilde R_{a\mu
\nu }\bigr)^{NGI}-\tilde C_{a0\nu }\eqno(19) 
$$
and $\bigl(i/F_\pi \cdot q^\mu \tilde R_{a\mu \nu }\bigr)^{NGI,GI}$
are given by the expressions (13) and (17) respectively.

Thus, the gauge invariance and the PCAC hypothesis allow to present the
amplitude of the electroproduction pion off nucleon as sum two term. First
of them represents a sum of the Born-contributions from the single-particle
intermediate states, the contributions from the matrix elements of the
commutator between the axial-vector current and electromagnetic current,
also the gauge noninvariant part of the electroproduction amplitude found
from the equation (11). It is expressed through the matrix elements of the
vector and the axial-vector currents and in the and it is expressed through
the form-factors vector and axial-vector currents. Consequently, this part of
the amplitude of the electroproduction is completely determinated in this
approach. The second term which is determined by formula (17) contains an
unknown non-pole part connecting with the axial-vector amplitude,
contributions from the many-particale intermediate states, the contribution
from the matrix element of the commutator between the divergence of the
axial-vector current and the electromagnetic current and represents part of
the amplitude not determined by basic hypotheses.

From the definition (14) it is followed that

\begin{center}
$$
\lim _{q_\mu \rightarrow 0}(i/F_\pi \cdot q^\mu \widetilde{R}_{a\mu \nu
})^{GI}=-\lim _{q_\mu \rightarrow 0}(i/F_\pi \cdot q^\mu \widetilde{R}_{a\mu
\nu })^{NGI}\eqno(20)%
$$
\end{center}

Therefore, from (18) we have

\begin{center}
$\lim _{q_\mu \rightarrow 0}(S_{a\nu }-S_{a\nu }(P))=-\lim _{q_\mu
\rightarrow 0}\widetilde{C}_{a0\nu }$
\end{center}

This corresponds to the presentation of the amplitude in the form

\begin{center}
$$
S_{a\nu }=S_{a\nu }(P)-\lim _{q_\mu \rightarrow 0}\widetilde{C}_{a0\nu
}+O(q),\eqno(21)%
$$
\end{center}

where the first two terms are model independent and give the result of the
standard $LET$. The term $O(q)$is the model dependent part of the amplitude.

Up to now amplitude $S_{a\nu }$ defined by the equation (18) is exact, as $%
S_{a\nu }(P)$ is the contribution from the all allowed single-particle
intermediate states. Near threshold one usually supposes that the $S_{a\nu
}(P)$ is the sum that includes only the contributions from the low lying
intermediate states. The gauge ivariance of the amplitude gives the
condition on the sums $S_{a\nu }(P)$ and $T_{a\mu }(P)$

\begin{center}
$$
F_\pi k^\nu S_{a\nu }(P)+q^\mu T_{a\mu }(P)+(q-k)^\nu \widetilde{C}_{a\nu }=0,%
\eqno(22)%
$$
\end{center}

where

\begin{center}
$\widetilde{C}_{a\nu }=ie\varepsilon _{a3c}\langle p_2\mid \widetilde{A}%
_{c\nu }(0)\mid p_1\rangle, $
\end{center}
if we use the chiral $SU(2)*SU(2)$ algebra commutation relations.

Therefore, the sums $S_{a\nu }(P)$ and $T_{a\mu }(P)$ contained the finite
numbers of the single-particle intermediate states must also satisfy the
condition (22).

Consider the $S_{a\nu }(P)$ and $T_{a\mu }(P)$ for the finite numbers of the
barion intermediate states with the spin $1/2$. Then if we take to account
the identities [18]

\begin{center}
$iq^\mu \langle p_2\mid \widetilde{A}_{a\mu }(0)\mid p_B\rangle =F_\pi
\langle p_2\mid j_{\pi a}(0)\mid p_1\rangle$ \\
\end{center}

$$
%\begin{center}
\hspace{50mm}-iN_{p_2}N_{p_B}\overline{u}(\overrightarrow{p}_2)\gamma _5(i\widehat{p}_B+m_B)%
\frac{\tau _a}2u(\overrightarrow{p}_B), %\end{center}
$$
$$
%\begin{center}
ik^\nu \langle p_B\mid j_\nu (0)\mid p_1\rangle =ieN_{p_B}N_{p_1}\overline{u}%
(\overrightarrow{p}_B)(i\widehat{p}_B+m_B)\frac{1+\tau _3}2u(\overrightarrow{p}_2) %\end{center}
$$
and the analogous identities for the nucleon matrix element $i\langle
p_{B^{^{\prime }}}\mid \widetilde{A}_{a\mu }(0)\mid p_1\rangle $ and
$\langle p_2\mid j_\nu (0)\mid p_{B^{\prime }}\rangle $, we have

\begin{center}
$
\begin{array}{c}
F_\pi k^\nu S_{a\nu }(B)+q^\mu T_{a\mu }(B)=0 \\ 
. 
\end{array}
$
\end{center}

The analogous result take place for the barion intermediate states with spin 
$\frac 32.$

Consider now the meson intermediate states($\pi -$meson, axial-vector and
vector mesons). If we take to account the identities

\begin{center}
$ik^\nu \langle \pi aq\mid j_\nu (0)\mid p_M\alpha _M\rangle =e\varepsilon
_{a3c}N_q(p_M^2+M^2)\langle 0\mid \varphi _c(0)\mid p_M\alpha _M\rangle $
\end{center}

and

\begin{center}
$\langle 0\mid \varphi _c\left( 0\right) \mid p_M\alpha _M\rangle =0,$
\end{center}

we have for the vector and axial-vector mesons

\begin{center}
$F_\pi k^\nu S_{a\nu }(M)=-e\varepsilon _{a3c}F_\pi \langle p_2\mid j_{\pi
c}\left( 0\right) \mid p_1\rangle .$
\end{center}

Besides, we have identity for the matrix element of the axial-vector current

\begin{center}
$-i(p_b-p_a)\langle p_b\mid \widetilde{A}_{c\mu }\left( 0\right) \mid
p_a\rangle =F_\pi \langle p_b\mid j_{\pi c}\left( 0\right) \mid p_a\rangle $
\end{center}

and as defined

\begin{center}
$\langle 0\mid \widetilde{A}_{a\mu }\left( 0\right) \mid \pi qb\rangle =0.$
\end{center}

Hence we have

\begin{center}
$(q-k)^\nu \widetilde{C}_{a\nu }=e\varepsilon _{a3c}F_\pi \langle p_2\mid
j_{\pi c}\left( 0\right) \mid p_1\rangle .$
\end{center}

As result

\begin{center}
$F_\pi k^\nu S_{a\nu }\left( M\right) +(q-k)^\nu \widetilde{C}_{a\nu }=0$
\end{center}

for the $\pi $-meson and finite numbers of the vector and axial-vector meson
states.

Thus the condition of the gauge invariance (22) is fulfilled for the finite
number barion states with the spin $\frac 12$ and $\frac 32$, for $\pi -$%
meson states and the finite numbers of the vector and axial-vector meson
states.

If we make use of the assumption that the sums in $S_{a\nu }(P)$ and $%
T_{a\mu }(P)$ include only the contributions from the nucleons and $\pi $
-meson, the amplitude $S_{a\nu }^0$ corresponds to the gauge invariant
amplitude of the electroproduction of pions off nucleons that is the sum of
the contribution from the nucleon exchange in the PV-theory, $\pi $ -meson
exchange and the contribution from the seagull term.

If in addition to the previous case, the $\Delta $ - resonance, the vector
and axial-vector mesons in the sums over intermediate states are included
and the chiral $SU(2)*SU(2)$ algebra commutation relations are used, the
amplitude $S_{a\nu }^0$ corresponds to the result of the calculation of the
amplitude of the electroproduction of pions off nucleons that were obtained
with the help of the chiral invariant Lagrangian [19] or the current algebra
model with hard pions [20]. The expression of this amplitude is given in
Refs [18] and [20]. Let us notice that in the general case the amplitude $%
S_{a\nu }^0$ contains the matrix element of the Schwinger term that is not
known. However, because we consider only the electroproduction of pions off
nucleons near the threshold, the contribution from the Schwinger term is not
considered.

As in the Ref. 9 we shall neglect the contributions from the many-particle
itermediate states, as the electroproduction of pions off nucleons is
considered near the threshold.

The amplitude connected with $"\Sigma -term"$ was calculated in the
Cloudy-Bag model. The result depends on the bag radius and the masses u and
d quarks. It can be submitted in the kind [9] 
$$
\frac{iq^\mu \delta _{\mu 0}}{F_\pi m_\pi ^2}\widetilde{\sum }_{a\nu }\cdot
\varepsilon ^\nu =-iN_{p_1}N_{p_2}\alpha \chi _f^{+}(f_{1a}i\overrightarrow{%
\sigma }\cdot \overrightarrow{\varepsilon }+f_{3a}i(\overrightarrow{\sigma }%
\cdot \widehat{\overrightarrow{k)}}(\widehat{\overrightarrow{q}}\cdot 
\overrightarrow{\varepsilon })+f_{4a}i(\overrightarrow{\sigma }\cdot 
\widehat{\overrightarrow{q}})\\ 
$$
$$
\hspace{30mm}(\widehat{\overrightarrow{q}}\cdot \overrightarrow{\varepsilon }%
)+f_{5a}i(\overrightarrow{\sigma }\cdot \widehat{\overrightarrow{q}})(%
\widehat{\overrightarrow{k}}\cdot \overrightarrow{\varepsilon })+f_{6a}i(%
\overrightarrow{\sigma }\cdot \widehat{\overrightarrow{k}})(\widehat{%
\overrightarrow{k}}\cdot \overrightarrow{\varepsilon }))\chi _i,\eqno(23) 
$$

$$
\begin{array}{c}
\\ 
f_1^{(+)}=A+\delta A, 
f_3^{(+)}=\left| \overrightarrow{q}\right| \left| \overrightarrow{k}%
\right| B,f_4^{(+)}=-\left| \overrightarrow{q}\right| ^2B, \\ 
f_5^{(+)}=-\left| 
\overrightarrow{k}\right| ^2B,f_6^{(+)}=\left| \overrightarrow{q}%
\right| \left| \overrightarrow{k}\right| B, \\ 
f_i^{(0)}=5/9f_i^{(+)},f_i^{(-)}=0. 
\end{array}
$$

The values $\alpha $, $A$, $B$ depend on the radius of the bag and the
masses of $u$ and $d$ quarks and are given in appendix 1. The corrections $%
\delta A^{(+,0}$ are connected with the amendments to MIT-Bag of model.
They arise in the Cloudy-Bag model to the first nontrivial order on the
meson field $\phi /F_\pi $.

We shall allocate the gauge-invariant part the from received expression ,
decomposing it on the linear - independent basis $O_ { i=1,...,6, \nu }, N_
{ 4, 7, \nu } $. As a result we have

$$
(\frac{iq^\mu \delta _{\mu 0}}{F_\pi m_\pi ^2}\widetilde{\sum }_{a\nu }\cdot
\varepsilon ^\nu )^{GI}=-iN_{p_1}N_{p_2}\alpha \chi _f^{+}(\sum_{i=1}^4f_{ia}%
\overrightarrow{V_i}\cdot \overrightarrow{\varepsilon }-(f_{1a}+\widehat{%
\overrightarrow{k}}\cdot \widetilde{\overrightarrow{q}}f_{3a}+f_{5a})\frac{%
k_0}{k^2}V_5\\ 
$$
$$
\hspace{55mm}-(\widehat{\overrightarrow{k}}\cdot \widehat{\overrightarrow{q}}%
f_{4a}+f_{6a})\frac{k_0}{k^2}V_6)\chi _i, \eqno(24) 
$$

where $V_{i=1,...,4}$ is the nonrelativistic structures entered by Adler [16
], 
$$
\begin{array}{c}
V_5=i 
\overrightarrow{\sigma }\cdot \widehat{\overrightarrow{k}}\widehat{%
\overrightarrow{k}}\cdot (k_0\overrightarrow{\varepsilon }-\varepsilon _0%
\overrightarrow{k}), \\ V_6=i\overrightarrow{\sigma }\cdot \widehat{%
\overrightarrow{q}}\widehat{\overrightarrow{k}}\cdot (k_0\overrightarrow{%
\varepsilon }-\varepsilon _0\overrightarrow{k}). 
\end{array}
$$

We shall allocate the GI-part from the amplitude of the electroproduction of
neutral pions off the protons connected with the non-pole part of the
axial-vector amplitude using the linear - independent basis $%
O_{i=1,...,6,\nu },N_{4,7,\nu }$. The result can be recorded in a kind 
$$
(\frac{iq^\mu }{F_\pi m_\pi ^2}\widetilde{T}_{a\mu \nu }\cdot \varepsilon
^\nu )^{GI}(NP)=-iN_{p_1}N_{p_2}\alpha \chi _f^{+}(\sum_{i=1}^4g_{ia}%
\overrightarrow{V_i}\cdot \overrightarrow{\varepsilon } 
$$
$$
\hspace{55mm}+[-k_0(g_{1a}+\widehat{\overrightarrow{k}}\cdot \widehat{%
\overrightarrow{q}}g_{3a}+g_{5a})+\left| \overrightarrow{k}\right|
g_{7a}]\frac 1{k^2}V_5 
$$
$$
\hspace{55mm}+[-k_0(\widehat{\overrightarrow{k}}\cdot \widehat{%
\overrightarrow{q}}g_{4a}+g_{6a})+\left| \overrightarrow{k}\right|
g_{8a}]\frac 1{k^2}V_6)\chi _i.\eqno(25) 
$$
Unknown functions $g_{i=1,...,8}$ are the factors of the decomposition of the
non-pole part of the axial-vector amplitude $({\ iq^\mu }/{\ F_\pi m_\pi ^2})%
\widetilde{T}_{a\mu \nu }\cdot \varepsilon ^\nu $ on the independent
structures $m_i$ of Walecka and Zucker [21]. Herein after we shall accept the
elementary model for unknown value $(({\ iq^\mu }/{\ F_\pi m_\pi ^2})%
\widetilde{T}_{a\mu \nu }\cdot \varepsilon ^\nu )^{GI}(NP)$ . Near to a
threshold we shall put $g_{i=1,...,4}=0$. It results in that $(({\ iq^\mu }/{%
\ F_\pi m_\pi ^2}) \linebreak \widetilde{T}_{a\mu \nu }\cdot \varepsilon ^\nu )^{GI}(NP)$
does not give the contribution to the transverse part of the amplitude of
the electroproduction of neutral pions off the protons near threshold. In
the limit $k^2=0$ it will be agreed with the assumption concerning the value 
$(({\ iq^\mu }/{\ F_{\pi m_\pi ^2}})\widetilde{T}_{a\mu \nu }\cdot
\varepsilon ^\nu )^{GI}(NP)$ in case of the photoproduct of neutral pions on
off protons accepted earlier in [9]. As according to the principle of the
analyticity the amplitude has only those singularities on the invariant
variables, which follow from the condition of the unitarity. Therefore the
remaining longitudinal part of the amplitude $(({\ iq^\mu }/{\ F_\pi m_\pi ^2%
})\widetilde{T}_{a\mu \nu }\cdot \varepsilon ^\nu )^{GI}(NP)$ should be
chosen such, to reduce the kinematic singularity on $k^2$ in the
amplitude $(({\ iq^\mu \delta _{\mu 0}}/{\ F_\pi m_\pi ^2})\widetilde{\sum }
_{a\nu }\cdot \varepsilon ^\nu )^{GI}$ wich arise under the decomposition on
the particularly chosen basis. The other part of the value $(({\ iq^\mu }/{\
F_\pi m_\pi ^2})\widetilde{T}_{a\mu \nu }\cdot \varepsilon ^\nu )^{GI}(NP)$
is chosen as the simple expression having one free parameter. Therefore near
threshold we shall put 
$$
\begin{array}{c}
-k_0g_{5a}+\left| 
\overrightarrow{k}\right| g_{7a}=-\alpha k_0(f_{1a}+\widehat{\overrightarrow{%
k}}\cdot \widehat{\overrightarrow{q}}f_{3a}+f_{5a})(-1+\beta k^2), \\ 
-k_0g_{6a}+\left| \overrightarrow{k}\right| g_{8a}=-\alpha k_0(\widehat{%
\overrightarrow{k}}\cdot \widehat{\overrightarrow{q}}f_{4a}+f_{6a})(-1+\beta
k^2),
\end{array}
$$
where $\beta $ is the free parameter. As a result we shall receive 
$$
(\frac{iq^\mu }{F_\pi m_\pi ^2}\widetilde{R}_{a\mu \nu }\cdot \varepsilon
^\nu )^{GI}=-iN_{p_1}N_{p_2}\alpha \chi _f^{+}(\sum_{i=1}^4f_{ia}%
\overrightarrow{V_i}\cdot \overrightarrow{\varepsilon }-k_0(f_{1a}+\widehat{%
\overrightarrow{k}}\cdot \widehat{\overrightarrow{q}}f_{3a}+f_{5a})\beta V_5 
$$
$$
\hspace{55mm}-k_0(\widehat{\overrightarrow{k}}\cdot \widehat{\overrightarrow{%
q}}f_{4a}+f_{6a})\beta V_6)\chi _i.\eqno(26)
$$
Thus, the amplitude of the electroproduction of
neutral pions off the protons contains only one free parameter $\beta $,
which enters in the longitudinal part of amplitude.

It should notice, that the allocation of the GI-part from amplitude is not a
simple procedure. As it is known, any amplitude can be presented as a sum of
the gauge-invariant and gauge-noninvariant parts by various ways. To realize
this, it is necessary to expand the amplitude on the various linear -
independent bases. For example, if instead of basis (12) one use the basis 
(12) in
which $N_{4\nu }$ $N_{7\nu }$ are replaed by $N_{3\nu 
}$ $N_{5\nu }$ [17], it results in the representation of the
amplitude in a kind similar (14). However the values $(i/F_\pi q^\mu \widetilde{R%
}_{a\mu \nu })^{^{\prime }GI,NGI}$ differ from appropriate values in initial
basis on gauge-invariant value
$$
R_{a\nu }^{GI}=\frac{2\alpha _a}{qk}\cdot \frac 1{k^2}O_{5\nu }+\frac{\beta
_a}{k^2}O_{6\nu }. 
$$
The physical results however should not depend on a way division of the
amplitude. This circumstance is necessary to take into account, when we
shall accept the certain model for unknown gauge-invariant part of the
amplitude connected with the non-pole part of the axial-vector amplitude.
The model assumptions for the gauge-invariant value ${\ (( {\ iq^\mu } / {%
\ F_\pi m_\pi ^2 } ) \widetilde { T } _ { a\mu \nu } \cdot \varepsilon ^\nu
) ^ { GI }( NP ) } $ should be mutually agreed in the various bases, so
that the physical results do not depend from choice of basis.

Thus, in the framework of the hypothesis of the gauge invariance and PCAC,
the dynamical assumption about the non-pole part of the matrix element of
the T-product of the axial-vector and the electromagnetic current and with ''%
$\Sigma -term$'' evaluated in the Cloudy-Bag model the amplitude of the
electroproduction of pions off nucleons near the threshold can be expressed
by formula (18), in which $S_{a\nu }^0$ is given by (19) and $({iq^\mu 
}/{F_\pi m_\pi ^2}\widetilde{R}_{a\mu \nu }\cdot \varepsilon ^\nu )^{GI}$ is
defined by (26). Near the threshold we shell consider the sum $S_{a\nu }(P)$
that contains only the contribution from the nucleons, $\Delta (1236)$%
-resonance, $\pi $-meson, $\rho ,\omega ,\phi $-vector mesons, A1-axial
vector exchanges. The contributions in the GI-invariant part of the
amplitude from them are given the appendix 2.

\noindent{\bf 3. The results and conclusion.} \vspace{2mm}

The cross-section of $(e,e^{\prime }\pi ^0)$ reaction can be written as

$$
d^3\sigma /dE_{e^{\prime }}d\Omega _{e^{\prime }}d\Omega _\pi ^{*}=\Gamma
_Vd\sigma /d\Omega _\pi ^{*}, 
$$

where
$$
\Gamma _V=\alpha /(2\pi ^2)\cdot k_{20}k_L/(k_{10}k^2)\cdot (1-\epsilon ) 
$$
is the flux of the virtual photons, $k_L=(W^2-m_p^2)/2m_p,W$ is the total
energy , $\epsilon $ is the polarization of the virtual photon,
$d\sigma /d\Omega_\pi ^{*}$ is the cross-section 
in the c.m. frame of the neutral pion
photo-production off protons by the virtual photons. Near threshold the
cross-section $d\sigma /d\Omega _\pi ^{*}$ can be written in the following
form

$$
\begin{array}{c}
d\sigma /d\Omega _\pi ^{*}=p_\pi ^{*}/q_L\cdot W/m_p\cdot \{A+
B\cos \theta _\pi^{*}+C\sin \theta _\pi ^{*}\cos \phi _\pi ^{*}+ \\ 
D\cos ^2\theta _\pi ^{*}+E\sin ^2\theta _\pi ^{*}\cos (2\phi _\pi ^{*})\},
\\  
\\  
\end{array}
$$
where p$_\pi ^{*}$ is the c.m. momentum of the pion, $\theta _\pi ^{*}$ is
the c.m. polar angle of the pion with respect to the transferred three
momentum $\overrightarrow{k}, \phi _\pi ^{*}$ is the c. m. azimutal angle of
the pion with respect to $\overrightarrow{q.}$ In the near threshold the
coefficients A, B, C, D, E can be written only in terms of the s- and p-wave
multipoles [22 ]. The s- and p- wave multipoles are expressed thorough the
amplitudes f$_i$ which are defined with help formula (18). In the expression
(18) the first term is the prediction of the current algebra with hard pions
, the second term is given by formula (26).

The summary amplitudes $ReE_{o+}(L_{0+})$ derived for the equal pion and
nucleon masses is to be applied at the degenerate threshold and above. In
reality, the $\pi ^0p$ and the $\pi ^{+}n$ thresholds are not degenerate. In
this case we shall describe the real process of the electroproduction of
neutral pions off protons at the $\pi ^{+}n$ threshold and above with the
help of the summery
amplitude (18) with modified masses and coupling constants. But when the
charged and neutral pion masses are different, a cusp occurs at the charged
pion threshold. As in the case of the neutral pion photoproduction [9] we
shall use the simple K- matrix calculations to into account this effect. We
shell treat the $E_{0+}(L_{0+})$ amplitude representing the current algebra
prediction plus contribution from the $"\Sigma -term"$ as the un-unitarized
input in a K- matrix formalism in order to find the change in the $%
ReE_{0+}(L_{0+})$ amplitudes from the $\pi ^0p$ threshold to $\pi ^{+}n$
theshold. As result we have the amplitudes between $\pi ^0p$ and $\pi ^{+}n$
thresholds as 
$$
Re[e(l)_{0+}(\pi ^0p)]=\overline{ReE}(\overline{ReL})_{0+}(\pi
^0p)-(4m_n^2m_{\pi ^{+}}^2-(W^2-m_n^2-m_{\pi ^{+}}^2)^2)^{1/2}/2W 
$$
$$
\hspace{55mm}*\sqrt{2}/3(a^{1/2}-a^{3/2})ReE(L)_{0+}^{LET}(\pi ^{+}n).%
\eqno(27) 
$$
Here $\overline{ReE}(\overline{ReL})_{0+}(\pi ^0p)$ is $ReE(L)_{0+}(\pi ^0p)$
amplitudes calculated for W between the $\pi ^0p$ and the $\pi ^{+}n$
thresholds with the help of the formula (18), in which the first term is
the result of
the current algebra with hard pions and second term is given by expression
(26 ); $ReE(L)_{0+}^{LET}(\pi ^{+}n)$ is the amplitude defined by LET for
the process $\gamma $ $p\rightarrow \pi ^{+}n$ on the $\pi ^{+}n$
threshold; $a^{1/2},a^{3/2}$ are the appropriate s-wave $\pi N$ -scattering
length.

In order to determine the unconstrained parameters $\beta $ we calculate the
differential cross-section of the neutral pion electroproduction off protons
for two values $\Delta W=W-m_p-m_{\pi ^0}=4$ and 8 MeV under $%
k^2=0.1(GeV/c)^2$ and $\epsilon =0.67$. The value of the parameter $\beta $
was taken as $\beta =1$. In the Fig. 1 the differential cross-section for $%
\phi _\pi ^0=0$ is shown by the solid curve, for $\phi _\pi ^0=180^0$ it is
shown by dashed curve. The results agrees with experimental data [11] and
close to the results of the CHPT model[14]. In the Fig. 2 the theoretical
calculation of the coefficients A, B, C are given for two model: 1) solid
curve is the calculations in our model; 2) dashed curve is calculations in
the current algebra with hard pions.

Having fixed parameter $\beta $, we can now make predictions for varies
lower photon virtualities. Later on we will consider s-wave cross-section, $%
ReE_{0+}$ and $ReL_{0+}$ near threshold.

In Fig. 3 the dependence of the s-wave cross-section from $k^2$ for $%
\epsilon $ =0.67 received in our model is shown, also the experimental data
are given in [10] for $\epsilon $= 0.58, 0.67, 0.79, in [11] for $\epsilon
$=0.67. It is shown also the dependence of s- wave cross-section from $k^2$
for $\epsilon $= 0.67 calculated with the help of the first term in the 
formula (18)
corresponding to the current algebra with hard pions. The current algebra
amplitude together with the additional contribution to the amplitude due to
the $"\Sigma -term"$ and the rescattering effect between the $\pi ^0p$ and the
$\pi ^{+}$ thresholds give the s-wave cross-section which are near the
experimental data in the $k^2$ range 0 - 0.12 $(GeV/c)^2$. The s-wave
cross-section dos not shown the flattening with increasing $k^2$. It dos not
consistent with the result of CHPT model. In our model the result of
calculating the s-wave cross-section confirms the significant contribution
from the $"\Sigma -term"$.

The amplitude received by us have been used also for the finding the
dependence of the s-wave multipoles as the function $\Delta W$ for various $%
k^2$. In Fig . 4 $ReE_{0+}(L_{0+})$ as function $\Delta W$ for $%
k^2=0.1(GeV/c)^2$ is given. The solid curve corresponds to the result received
with the help of the formules (18), (26) and (27). The dashed curve is the
current algebra result. It is shown that $ReE_{0+}$has changed sign as
compared to typical cusp effect at the opening of the $\pi ^{+}n$ threshold.
In contrast, $ReL_{0+}$ is essential energy-independent. It is shown also,
that the contribution $"\Sigma -term"$is important. The result for $ReL_{0+}$
is similar to result of CHPT model, however $ReE_{0+}$ is larger than in
CHPT model. In Fig. 3 ''experimental'' points [11] received with using
p-waves calculated theoretically in [23] are also shown. The model
dependence in the resulting values $ReE_{0+}(L_{0+})$ is estimated to be
about 10\%.

In Fig. 5 the energy dependence of the $ReE_{0+}(L_{0+})$ are given for the
small $k^2=0.6,0.5,0.4(GeV/c)^2$. The $ReE_{0+}$ increased with rise of $k^2$.
It is also seen the display of the ''cusp-effect''. Our $ReE_{0+}(L_{0+})$ for
the small $k^2$ are also large than the prediction of the CHPT model.

Thus, the base results are following: the phenomenological model for the
description of the neutral electroproduction pion off proton near threshold
are constructed; the predictions of the $a_0,ReE_{0+},ReL_{0+}$ do not
contradict to the available experimental data; it may to do the conclusion 
about the significant contribution from the $"\Sigma -term"$. The energy 
behavior of the s-wave transverse and longitudinal multipoles are different
from the predictions of CHPT in the same kinematical area. New experimental 
dates are nessery for the examination of the theoretical predictions.

%\noindent {\bf References.}

\noindent {\bf Figure captions.}\vspace{2mm}

Fig. 1. The differential cross-section for $k^2=0.1(GeV)^2$. The dashed and
solid curves are the calculations of the differential cross-section in our
model (CA predictions plus contribution from $"\Sigma -term"$ taking into
account the rescattering effect between the $\pi ^0p$ and the $\pi ^{+}n$
thresholds) for $\phi _\pi ^{*}=0$ and $\phi _\pi ^{*}=180^0$ respectively.
The experimental differential cross-sections are given for $-30^0\prec \phi
_\pi ^{*}\prec 30^0$ (solid squares) and for $150^0\prec \phi _\pi ^{*}\prec
210^0$ (solid circles) [11].

Fig. 2. The coefficients A, B, C for $k^2=0.1(GeV)^2$. The solid curve is the
result of the calculation in our model (CA prediction plus contribution from 
$"\Sigma -term"$ taking into account the rescattering effect between the $\pi
^0p $ and the $\pi ^{+}n$ thresholds). The dashed curve is the result of the
CA. Experimental points are taken from [11].

Fig. 3. The s-wave differential cross-section. The solid curve is the
calculation in our model for $\epsilon =0.67.$ The dashed curve is the
result of the CA with hard pions for $\epsilon =0.67.$ The experimental
points: solid squares, open squares, solid circles for $\epsilon
=0.58,0.62,0.79$ respectively [10]; solid triangles for $\epsilon =0.67$
[11].

Fig. 4. The energy dependence of the $ReE_{0+}$ and the $ReL_{0+}$
multipoles for $k^2=0.1(GeV/c)^2.$ The solid curve is the result of CA plus
contribution from $"\Sigma -term"$ taking into account the rescattering
effect between the $\pi ^0p$ and the $\pi ^{+}n$ thresholds. The dashed curve
is result of CA.

Fig. 5. The energy dependence of the $ReE_{0+}$ and the $ReL_{0+}$
multipoles for small $k^2.$ The solid, dashed and dotted curve are the
calculations in CA plus the contribution from $"\Sigma -term"$ taking into
account the rescattering effect between the $\pi ^0p$ and the $\pi ^{+}n$
thresholds for $k^2=0.04,0.05,0.06(GeV/c)^2.$

\noindent{\bf Appendix 1}\vspace{2mm}

The coefficient $\alpha $ is given in [9] as

$$
\alpha =-\frac e{F_\pi }\cdot 
\frac{mq^0}{m_\pi ^2}\cdot \frac{m_N}{4\pi (m_N+m_\pi )}\cdot \frac{%
(E_{N_1}E_{N_2}I(p_2))^{1/2}}{(m_N^2I(p_1))^{1/2}}\cdot z_Nz_1, \\  
$$

where $m=(m_u+m_d)/2$, $m_u$ and $m_d$ are the masses of the u and d quarks;
I(p)= $\int dr\Phi _N(\overrightarrow{r})e^{ipr}$ and $\Phi _N$ is the
overlap function of the nucleon; $z_{N}, z_1$ are the renormalization
constants. The functions A and B are given in [9] as

$$
A( 
\overrightarrow{k}-\overrightarrow{q})=N^2(\kappa )[E_{+}^2I_{00}(\alpha
,\beta )+E_{-}^2I_{01}(\alpha ,\beta )-2M_{11}(\alpha ,\beta )]
$$

$$
 B(\overrightarrow{k}-\overrightarrow{q})=N^2(\kappa )\frac{2E_{-}^2}
{{\beta}^2}[I_{10}(\alpha ,\beta )-3M_{11}(\alpha ,\beta )], 
$$

where

$$
\begin{array}{c}
I_{nm}(\alpha ,\beta )=\int_0^{R_N}drr^2j_n^2(\alpha r)j_m(\beta r) \\ 
M_{nm}(\alpha ,\beta )=\int_0^{R_N}drr^2j_n^2(\alpha r)j_m(\beta r)/
\beta r, 
\end{array}
$$

$\alpha =\kappa /R_N,\beta =\mid \overrightarrow{k}-\overrightarrow{q}\mid
,N^{-2}(\kappa )$ is normalization constant,

$$
E_{\pm }=(\frac{E\pm m}E)^{1/2},E=\frac{[\kappa ^2+(mR_N)^2]^{1/2}}{R_N}, 
$$

$\kappa $ is determined by the linear boundery condition.

\noindent{\bf Appendix 2}\vspace{2mm} 

The invariant amplitudes from the nucleon exchange are

$$
^NA_i^{(\pm ,0)}=^NH_i^{(\pm ,0)}[^{}\frac{^N\varphi _i^{(\pm ,0)}(s,k^2)}{%
s-m_N^2}+^N\eta _i^{(\pm ,0)}\frac{^N\varphi _i^{(\pm ,0)}(u,k^2)}{u-m_N^2}%
]+ 
$$

$$
^NR_i^{(\pm ,0)}(s,k^2)\pm ^N\varsigma_i^{(\pm ,0)}\cdot^N R_i^{(\pm ,0)}
(u,k^2), 
$$

where 

$$
^N\varphi _i^{(\pm ,0)}(s,k^2)  =g_{\pi NN}(s)F_1^{V,S}(s,k^2),~~~~~ i=1,3,4
$$
$$
~~~~~~~~~~~~~ =\frac 1{qk}g_{\pi NN}(s)F_1^{V,S}(s,k^2),~~~  i=2 
$$
$$  
~~~~~~~~~~~~~~~~~ =0,~~~~~~~~~~~~~~~~~~~~~~~~~~~~~~ i=5,6 ,
$$

$$
^NR_i^{(\pm ,0)}(s,k^2)  =(g_{1\pi NN}(s)-\frac 1{2F_\pi
})F_2^{V,S}(s,k^2), ~~~ i=1  
$$
$$
~~~~~~~~~~~~~~~~~~~~~~~~ =0,~~~~~~~~~~~~~~~~~~~~~~~~~~~~~~~~~~~~~~~~ i=2,3,4 
$$
$$
~~~~~~~~~~~~~~~ =-\frac 1{k^2}\cdot \frac 1{qk}g_{\pi NN}(s)F_1^{V,S}(s,k^2),
~~~~~~~~ i=5  
$$
$$
~~~~~~~~~~~~~~~~~~~ =\frac 1{k^2}(g_{1\pi NN}(s)-\frac 1{2F_\pi })
F_1^{V,S}(s,k^2), ~~ i=6,
$$ 

$$
^NH_i^{(\pm ,0)} =\cases{1,& i=1; \cr
-1,&  i=2,3,4,\cr}
$$
$$
^N\eta _i^{(\pm ,0)} =\cases{\pm,& $i=1,2,4$;\cr
 \mp,&  $i=3$,\cr}
$$

$$
^N\varsigma _i^{(\pm ,0)}=\cases{\pm,& $i=1$; \cr
\mp, &$i=5,6$. \cr}
$$

Here $g_{\pi NN}(-m_N^2)=g_{\pi NN}$ is the constant of the pseudo-scalar
cupling, $g_{\pi NN}^2/4\pi =14.4$ and
$$ g_{1\pi NN}(-m_\pi ^2)=\frac1{2F_\pi}
-\frac{g_{\pi NN}}{2m_N};
$$
$F_i^{V,S}(-m_N^2,k^2)=F_i^{V,S}(k^2)
$
are the nucleon
isoscalar and isovector form factors [20] whith normalization%

$$
F_1^V(0)=F_2^S(0)=\frac e2, F_2^V(0)=\kappa ^V=1.85\frac
e{2m_N}, F_2^S(0)=\kappa ^S=-0.065\frac e{2m_N}. \\  
$$

The invariant amplitudes from the $\Delta $ resonance are 
$$
^\Delta A_i^{(\pm )}=^\Delta H^{(\pm )}(k^2)[\frac{^\Delta \varphi _i^{(\pm
)}(s,k^2)}{s-m_N^2}+^\Delta \eta _i^{(\pm )}\frac{^\Delta \varphi _i^{(\pm
)}(u,k^2)}{u-m_N^2}] ,
$$

where 
$$
^\Delta \eta_i^{(\pm)}=\cases{\pm,& $i=1,2,4$; \cr
\mp,& $i=3,5,6 $ \cr},%
$$
$$
^{\Delta}H^{(+)}(k^2)=- g_{\pi \Delta N}\frac{ek^{*}}{6m_N} \frac{m_{\rho} ^2%
}{m_{\rho}^2+k^2}, 
$$
$$
^{\Delta}H^{(-)}(k^2)= -1/2 \cdot ^{\Delta}H^{(+)}(k^2) ,
$$

$$
^\Delta \varphi _1^{(\pm )}(s,k^2)=-4(t+k^2)+m_NE(s)+2k^2Y(s) ,
$$
$$
^\Delta \varphi _2^{(\pm )}(s,k^2)=8-\frac{4k^2}{t-m_\pi ^2+k^2}Y(s) ,
$$
$$
^\Delta \varphi _3^{(\pm )}(s,k^2)=4(m_N+m_\Delta )-2m_NY(s) ,
$$
$$
^\Delta \varphi _4^{(\pm )}(s,k^2)=-4(m_N+m_\Delta )-2m_N^{}Y(s), 
$$
$$
^\Delta \varphi _5^{(\pm )}(s,k^2)=4\frac{m_N^2-s}{t-m_\pi ^2+k^2}Y(s) ,
$$
$$
^\Delta \varphi _6^{(\pm )}(s,k^2)=-E(s), 
$$
and 
$$
E(s)=\frac 4{3m_\Delta }[m_N^2-s-2m_\pi ^2+\frac{m_N}{m_\Delta }%
(m_N^2-s-m_\pi ^2)] ,
$$
$$
Y(s)=1+\frac{2m_N}{m_\Delta ^2}-\frac{2m_N^2-2m_\pi ^2-s}{3m_\Delta } .
$$
The constants 
$g_{\pi \Delta N}=0.290\pm 0.006$, $k^{*}=5.02$ [20].

The invariant amplitudes from the $\pi$- meson exchange are

$$
^\pi A_i^{(\pm ,0)}=\frac{^\pi \varphi _i^{(\pm ,0)}(t,k^2)}{t-m_\pi ^2} ,
$$
$$
\begin{array}{cll}
^\pi \varphi _i^{(\pm ,0)}(t,k^2) & =0, & i=1,2,3,4,6 \\ 
^\pi \varphi _i^{(\pm ,0)}(t,k^2) & =2/k^2\cdot g_{\pi NN}(t)F^\pi (t,k^2)
(\frac
12\mp \frac 12), & i=5. \\  
&  &  
\end{array}
$$
Here $F^\pi (-m_\pi ^2,k^2)=F^\pi (k^2)$ is the pion form factor with
normalization $F^\pi (0)=e$.

The invariant amplitudes from the $A_1$- meson exchange are 
$$
^{A_1}A_i^{(\pm ,0)}=\frac{^{A_1}\varphi _i^{(\pm ,0)}(t,k^2)}{t-m_{A_1}^2}%
+^{A_1}R_i^{(\pm ,0)}(k^2) ,
$$
$$
\begin{array}{cll}
^{A_1}\varphi _i^{(\pm ,0)}(s,k^2) & =0, & i=1,2,4 \\  
&  &  \\  
& =g_{\pi NN}/m_N\cdot (\lambda _A-1)F_1^{V,S}(k^2)(\frac 12\mp \frac 12), & 
i=3 \\  
&  &  \\  
& =g_{\pi NN}/{m_\rho ^2}\cdot (2-\lambda _A)F_1^{V,S}(k^2)(\frac 12\mp
\frac 12), & i=5 \\  
&  &  \\  
& =-g_{\pi NN}/{m_N}\cdot F_1^{V,S}(k^2)(\frac 12\mp \frac 12), & i=6, \\  
&  &  \\  
&  &  
\end{array} 
$$
$$
\begin{array}{cll}
^{A_1}R_i^{(\pm ,0)}(k^2) & =0, & i=1,2,3,4,5 \\  
&  &  \\ 
^{A_1}R_i^{(\pm ,0)}(k^2) & =-g_{\pi NN}/m_N\cdot \frac
1{k^2}F_1^{V,S}(\frac 12\mp \frac 12), & i=6,
\end{array}
$$
where $\lambda _A=0.4$ [20].

The invariant amplitude from the $\rho $-meson exchange are 
$$
^\rho A_i^{(\pm ,0)}=\frac{^\rho \varphi _i^{(\pm ,0)}(t,k^2)}{t-m_\rho ^2}, 
$$
$$
^\rho \varphi _i^{(\pm )}(t,k^2)=0,\hspace{5cm}i=1,2,3,4,6 
$$
$$
^\rho \varphi _i^{(0)}(t,k^2)=f_{\rho \pi \gamma }(k^2)g_{2\rho NN}t,
\hspace{4cm}i=1 
$$
$$
^\rho \varphi _i^{(0)}(t,k^2)=-f_{\rho \pi \gamma }(k^2)g_{\rho NN}\frac{%
t-m_\pi ^2-k^2}{t-m_\pi ^2+k^2},\hspace{2cm}i=2 
$$
$$
^\rho \varphi _i^{(0)}(t,k^2)=0,\hspace{6.8cm}i=3,6 
$$
$$
^\rho \varphi _i^{(0)}(t,k^2)=f_{\rho \pi \gamma }(k^2)g_{1\rho NN},%
\hspace{4.3cm}i=4 
$$
$$
^\rho \varphi _i^{(0)}(t,k^2)=f_{\rho \pi \gamma }(k^2)g_{2\rho NN}\frac{u-s%
}{t-m{\pi }^2+k^2},\hspace{2cm}i=5 .
$$

Here 
$$
f_{\rho \pi \gamma }=
\frac{4e}{\sqrt{3}}(\frac{g_{\varpi 8}g_{\rho \pi \varpi }}{m_\rho ^2+k^2}+%
\frac{g_{\varphi 8}g_{\rho \pi \varphi }}{m_\rho ^2+k^2}),
$$

$$ g_\rho
g_{1\rho NN}=\frac 12m_\rho ^2,g_\rho g_{2\rho NN}=-
\frac{\kappa ^V}{2m_N}m_\rho ^2,
$$

$$\\ g_{\varpi 8}=
\frac{m_\varpi }{m_\rho }g_\rho \sin \theta , g_{\varphi 8}=
\frac{m_\varphi }{m_\rho }g_\rho \cos \theta ,  
$$

and $g_{\rho \pi \omega }=-0.597m_\pi ^{-1}, g_{\rho \pi \varphi
}= - 0.0193m_\pi ^{-1}$.

The invariant amplitudes from the $\omega ,\varphi $-meson exchanges are 
$$
^{\omega ,\varphi }A_i^{(+0)}=[\frac{^\omega \varphi _i^{(+)}(t,k^2)}{%
t-m_\omega ^2}+\frac{^\varphi \varphi _i^{(+)}(t,k^2)}{t-m_\phi ^2}] ,
$$
$$
^{\omega ,\varphi }\varphi _i^{(+)}=f_{\omega ,\varphi \pi \gamma }(k^2)g_
{2\omega,\varphi NN}t,\hspace{5cm}i=1 
$$
$$
^{\omega ,\varphi }\varphi _i^{(+)}=f_{\omega ,\varphi \pi \gamma }(k^2)g_
{2\omega,\varphi NN}\frac{t-m_\pi ^2-k^2}{t-m_\pi ^2+k^2},\hspace{3cm}i=2 
$$
$$
^{\omega ,\varphi }\varphi _i^{(+)}=0,\hspace{8.5cm}i=3,6 
$$
$$
^{\omega ,\varphi }\varphi _i^{(+)}=f_{\omega ,\varphi \pi \gamma }(k^2)g_
{1\omega,\varphi NN},\hspace{5.6cm}i=4 
$$
$$
^{\omega ,\varphi }\varphi _i^{(+)}=[f_{\omega ,\varphi \pi \gamma
}(k^2)g_{2\omega ,\varphi NN}]\frac{u-s}{t-m{\pi }^2+k^2},\hspace{3cm}i=5, 
$$

where 
$$
f_{\varpi \pi \gamma }(k^2)=\frac{4eg_\rho }{m_\rho ^2+k^2}g_{\rho \pi
\omega },f_{\varphi \pi \gamma}(k^2)=\frac{4eg_\rho }{m_\rho ^2+k^2}%
g_{\rho \pi \varphi}. 
$$

The contribution from the current algebra term are 
$$
^{CA}A_i^{(\pm ,0)}=0,\hspace{5.0cm}i=1,...,4 
$$
$$
^{CA}A_i^{(\pm ,0)}=-\frac e{2F_\pi }\cdot \frac 1{k^2}\tilde g_P(t)(\frac
12\mp \frac 12),\hspace{2cm}i=5 
$$
$$
^{CA}A_i^{(\pm ,0)}=\frac e{2F_\pi }\cdot \frac 1{k^2}\tilde g_A(t)(\frac
12\mp \frac 12),\hspace{2.7cm}i=6, 
$$
where 
$$
\tilde g_A(t)=g_A(t)+\frac{F_\pi }{m_N}\cdot \frac{m_{A_1}^2g_{\pi NN}}
{t-m_{A_1}^2}, 
$$
$$
\tilde g_P(t)=g_P(t)+\frac{2F_\pi g}{t-m_\pi ^2}-\frac{2F_{\pi}g_{\pi NN}}
{t-m_{A_1}^2},
$$
$$
g_A=g_A(0)(1-\frac t{{\tilde m}_{A_1}^2})^{-2},{\tilde m}_{A_1}^2=(0.95\pm
0.09)GeV. 
$$

\newpage
\unitlength=1cm
\begin{picture}(18,30)
\put(-2,18){\epsfxsize=10cm\epsfbox{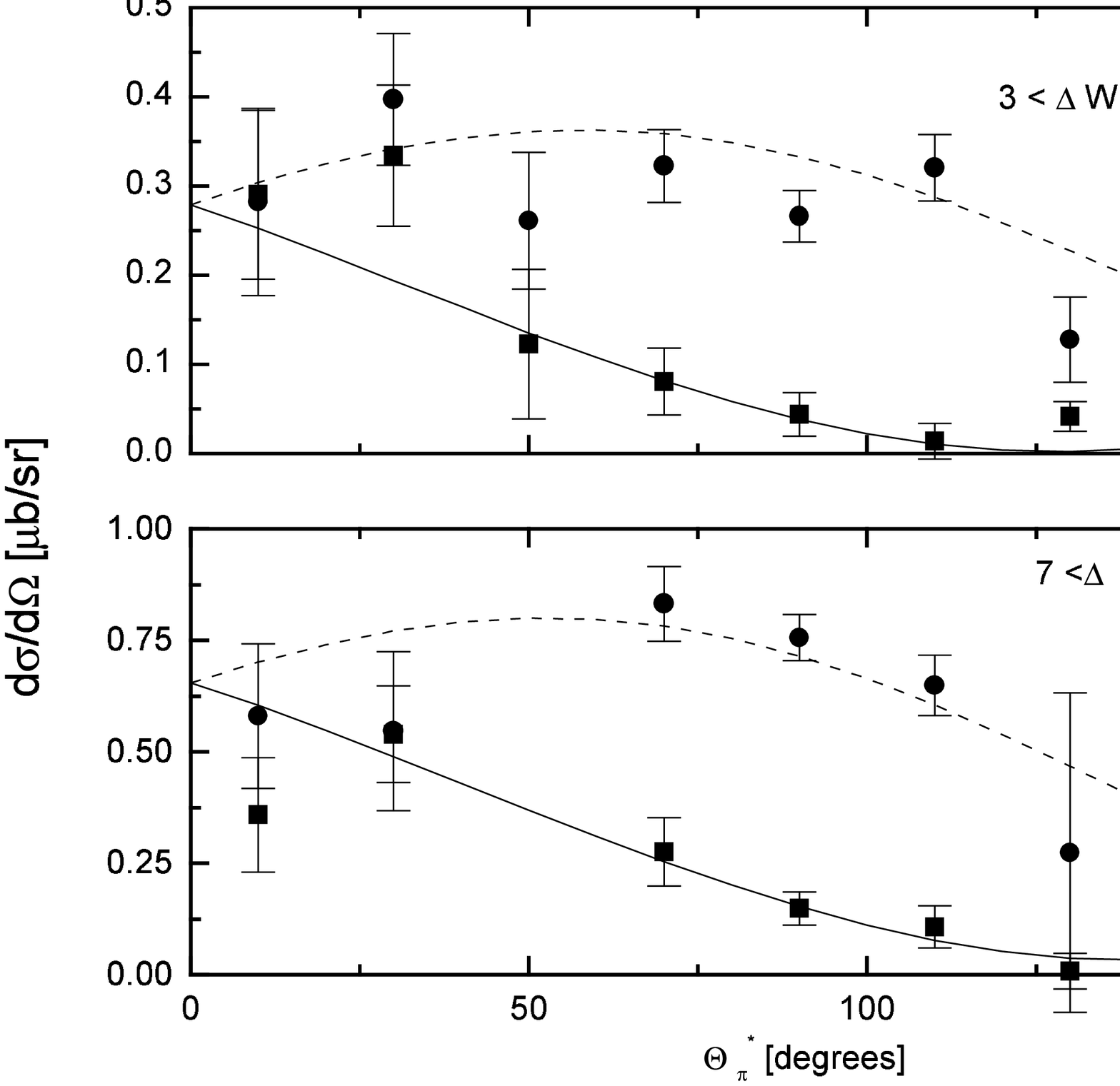}}
\put(-2,6.5){\epsfxsize=11cm\epsfbox{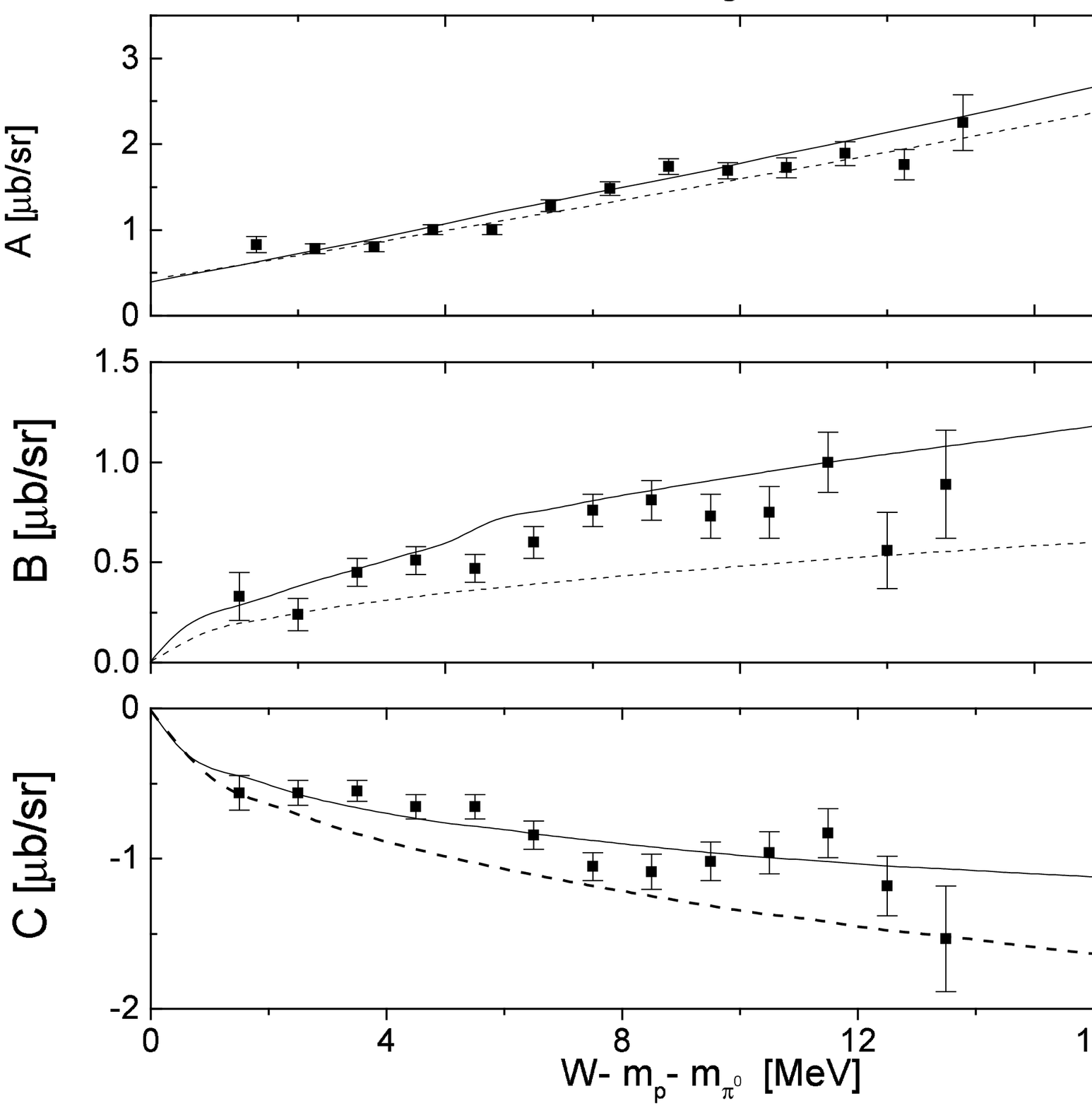}}
\end{picture}

\newpage
\unitlength=1cm
\begin{picture}(18,30)
\put(-2,17){\epsfxsize=10cm\epsfbox{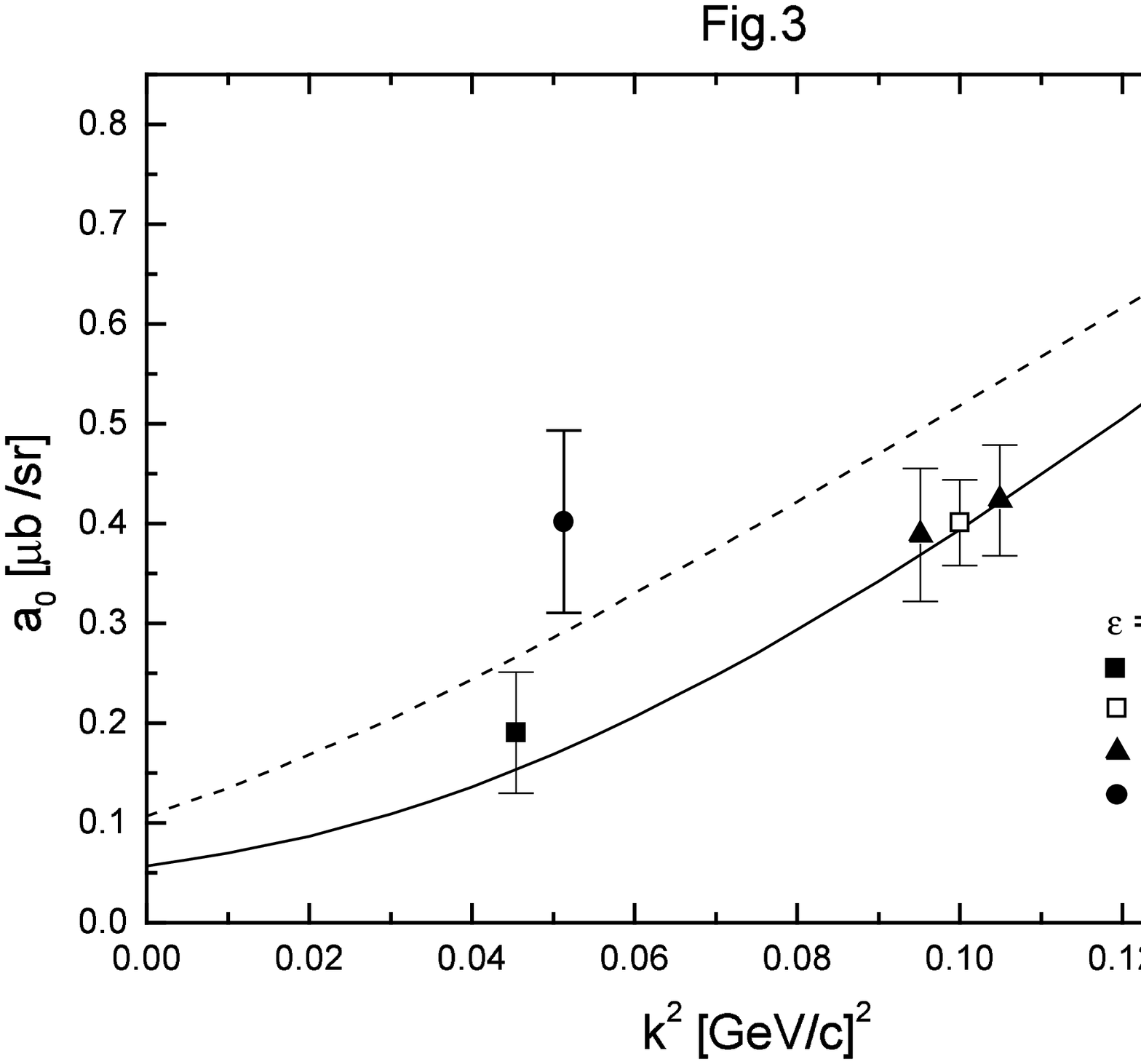}}
\put(-2,7){\epsfxsize=10cm\epsfbox{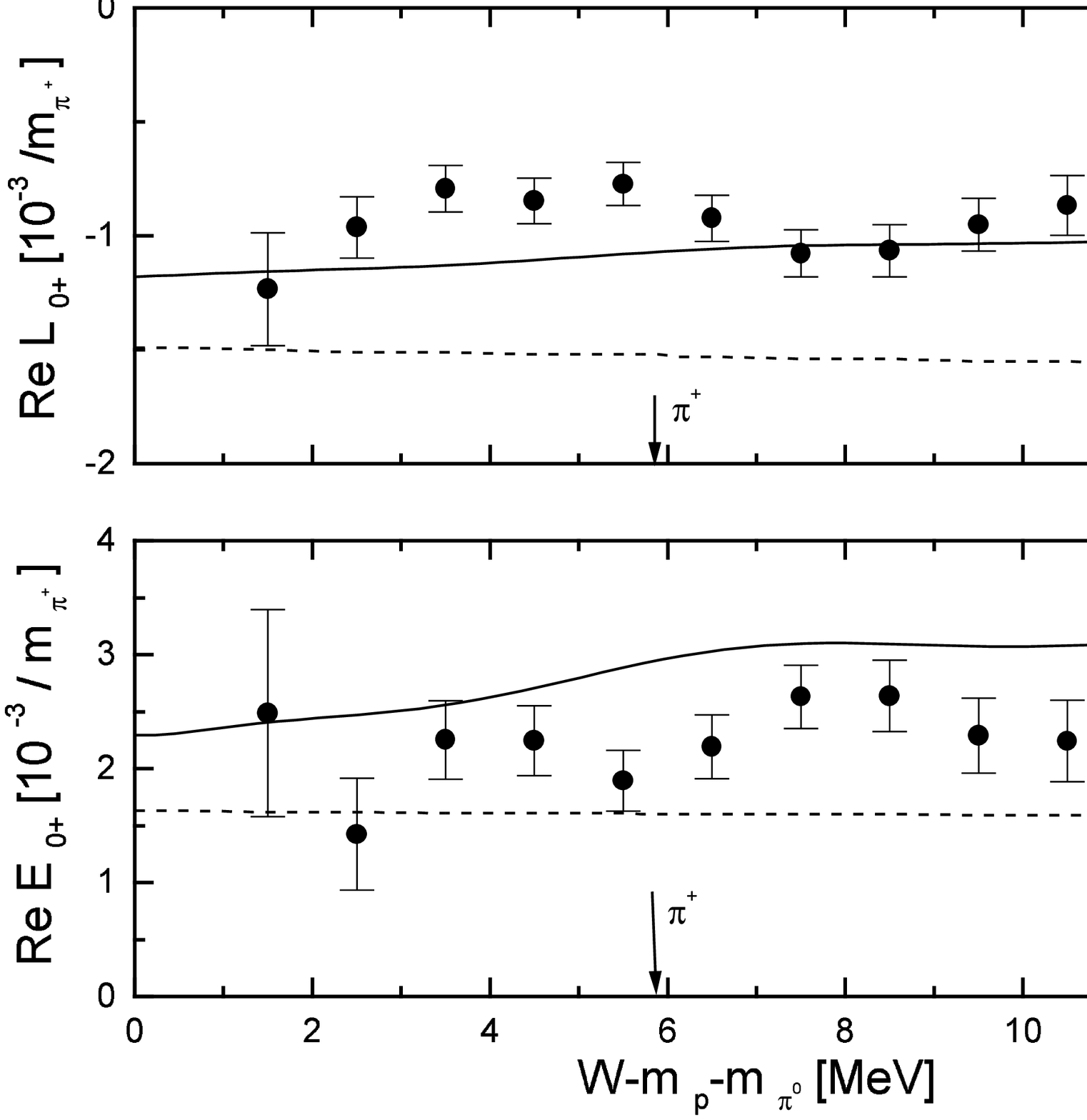}}
\end{picture}

\newpage
\unitlength=1cm
\begin{picture}(18,30)
\put(-2,15){\epsfxsize=10cm\epsfbox{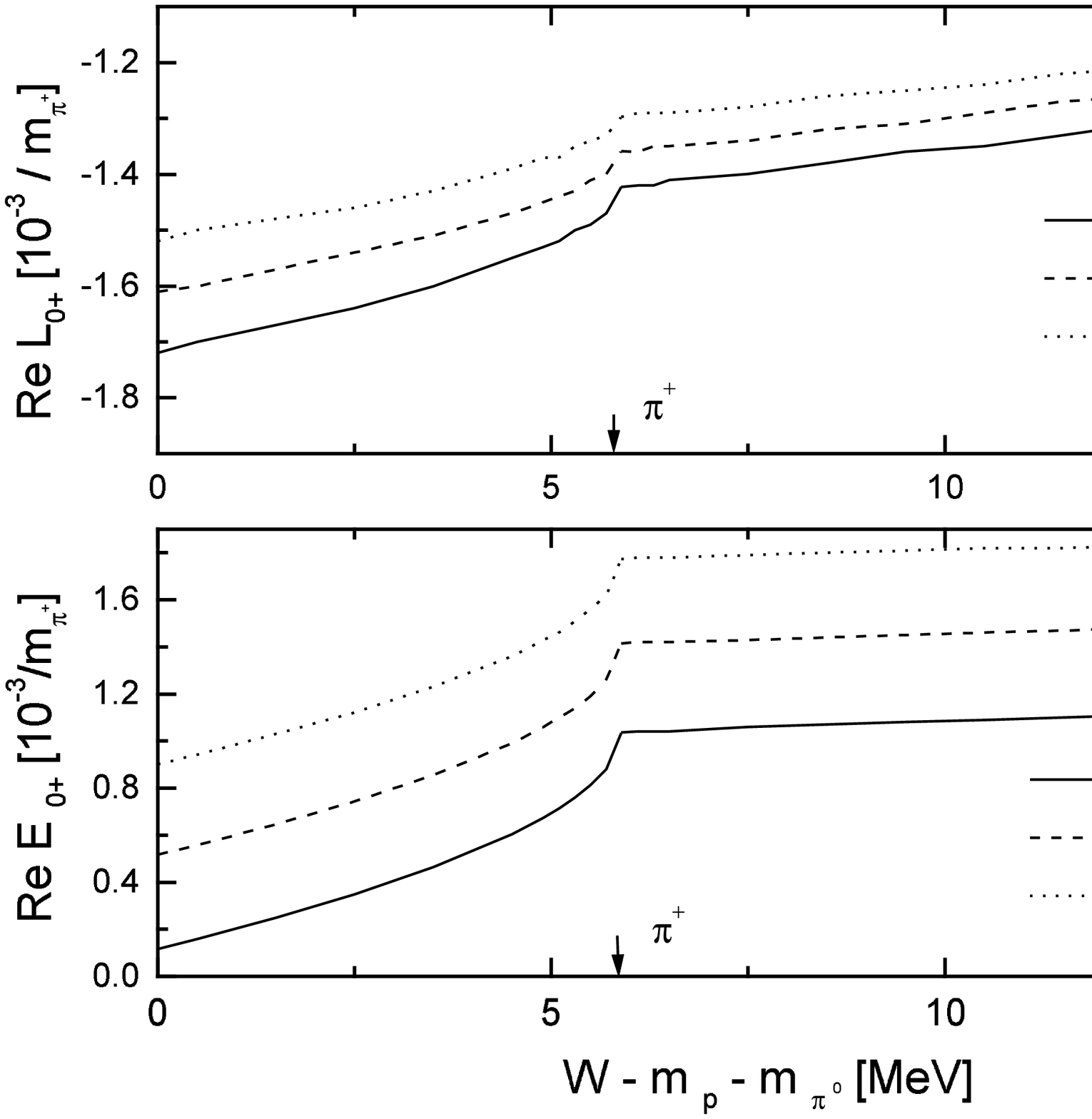}}
\end{picture}
\end{document}